\documentclass[reprint,aps,prc,twocolumn,amsmath,superscriptaddress,floatfix,nofootinbib]{revtex4-2}

\usepackage{setspace,ulem, epsfig,amssymb,amsfonts,amsmath,mathtools,bm,color,xcolor,graphicx,braket,adjustbox,esint,upgreek,verbatim,ctable,array,float,dcolumn}

\usepackage[colorlinks, linkcolor=red,anchorcolor=green,citecolor=blue]{hyperref}

\newcommand{\ignore}[1]{}

\begin{document}
%\begin{spacing}{1.0}
\title{Investigating $J/\psi$ spin alignment in heavy-ion collisions within a two-component transport model}

\author{Yida Yang}
\affiliation{Department of Physics, School of Science, Tianjin University, Tianjin 300354, China}

\author{Anping Huang}
\email{huanganping425@cumt.edu.cn}
\affiliation{School of Material Science and Physics, China University of Mining and Technology, Xuzhou 221116, China}

\author{Baoyi Chen}
\email{baoyi.chen@tju.edu.cn}
\affiliation{Department of Physics, School of Science, Tianjin University, Tianjin 300354, China}

\date{\today}

\begin{abstract}
We investigate the spin alignment of $J/\psi$ mesons in relativistic heavy-ion collisions within a two-component Boltzmann transport model. Starting from the relativistic spin Boltzmann equation, we derive the spin density matrix element $\rho_{00}$ under a non-relativistic approximation for heavy quarks. To interpret the recent ALICE measurements in Pb+Pb collisions, the observed $\rho_{00}$ is described as a $p_T$-dependent mixture of contributions from primordial production and the coalescence process. At forward rapidity, the $p_T$ dependence of charmonium $\rho_{00}$ is well reproduced by this two-component mechanism: at low $p_T$, charmonium production is dominated by the coalescence of partially polarized charm quarks induced by thermal vorticity; with increasing $p_T$, primordially produced charmonia become dominant, causing $\rho_{00}$ to approach $1/3$. To further test this spin alignment mechanism, we provide predictions for the $J/\psi$ $\rho_{00}$ in the mid-rapidity region, which exhibits a distinct $p_T$ trend due to the kinematic suppression of the thermal vorticity contribution. This study elucidates the underlying mechanism of $J/\psi$ spin alignment and advances our understanding of heavy quarkonium spin dynamics in strongly interacting matter.

\end{abstract}

\maketitle

\section{Introduction}
In non-central relativistic heavy-ion collisions, a huge initial orbital angular momentum (OAM) perpendicular to the reaction plane is carried by the system.  Part of the OAM is converted into the thermal vorticity of the quark-gluon plasma (QGP) \cite{dengVorticityHeavyionCollisions2016,weiThermalVorticitySpin2019,Huang2021,becattiniStudyVorticityFormation2015}. Simultaneously, extremely strong electromagnetic fields are generated in non-central collisions \cite{dengEventbyeventGenerationElectromagnetic2012,skokovEstimateMagneticField2009,tuchinParticleProductionStrong2013,voronyukElectromagneticFieldEvolution2011a}.  These phenomena can induce the global spin polarization of partons (e.g., quarks) within the QGP \cite{buzzegoliSpinPolarizationInduced2023,becattiniRelativisticDistributionFunction2013,becattiniGlobalHyperonPolarization2017,becattiniPolarizationVorticityQuark2020,Betz:2007kg,fangPolarizationMassiveFermions2016}. Although the spin polarization of quarks cannot be directly observed due to the confinement nature \cite{wilsonConfinementQuarks1974} of QCD, the spin polarization or spin alignment of hadrons formed after the hadronization process can be observed. This consequently helps us understand the mechanisms underlying quark polarization.

The global spin polarization of hyperons, such as $\Lambda$, $\bar{\Lambda}$, $\Xi$, and $\Omega$, has been observed by the STAR Collaboration in Au+Au collisions \cite{thestarcollaborationGlobalHyperonPolarization2017,adamGlobalPolarizationHyperons2018,adamGlobalPolarizationHyperons2021}. Since both hyperons and quarks are fermions, the hyperon polarization is closely related to the polarization of individual constituent quarks \cite{shengWhatCanWe2020,liangGloballyPolarizedQuarkGluon2005}. These measurements strongly support the scenario where thermal vorticity induces quark polarization. Conversely, they place constraints on the expected contributions from strong magnetic fields. Beyond hyperons, the overall spin orientation of spin-1 vector mesons is of great interest. This observable is widely known as spin alignment. In the quark coalescence model for hadronization, the spin alignment of a vector meson is determined by the polarization of its two constituent quarks and their correlations \cite{liangSpinAlignmentVector2005,lvGlobalQuarkSpin2024,xuSpinAlignmentVector2024,shengImprovedQuarkCoalescence2020,yangQuarkCoalescenceModel2018,chenFluctuationsCorrelationsQuark2025}.

Unlike the spin polarization of fermions, the spin alignment of vector mesons is described by a spin density matrix. The most relevant matrix element for experimental observations is $\rho_{00}$. This element is typically extracted from the polar angle distribution of the meson's decay products \cite{liangSpinAlignmentVector2005,starcollaborationPatternGlobalSpin2023,yangQuarkCoalescenceModel2018}. Specifically, this distribution can be expressed as:
\begin{equation}
    \frac{\mathrm{d}N}{\mathrm{d}(\mathrm{cos}\theta^*)}\propto(1-\rho_{00})+(3\rho_{00}-1)\mathrm{cos}^2\theta^*,
\end{equation}
where $\theta^*$ is the polar angle between the spin quantization axis and the momentum direction of the decay particle. In 2008, the STAR Collaboration observed no significant spin alignment for $\phi$ mesons ($s\bar{s}$ state) in Au+Au collisions at $\sqrt{s_{NN}} = 200$ GeV \cite{abelevPartonicFlowMeson2007}. However, substantial $\phi$ meson spin alignment signals have been observed at lower collision energies recently \cite{starcollaborationPatternGlobalSpin2023}. Nevertheless, these experimental data for $\phi$ mesons cannot be fully explained by thermal vorticity alone. This discrepancy suggests the existence of other mechanisms influencing quark polarization and the subsequent vector meson spin alignment \cite{shengWhatCanWe2020,shengImprovedQuarkCoalescence2020,shengSpinAlignmentVector2023}.

To further explore the mechanisms driving quark polarization and vector meson spin alignment in the QGP, we focus on the $J/\psi$ meson. Unlike the $\phi$ meson, the $J/\psi$ meson is a heavy quarkonium state composed of a $c\bar{c}$ pair. It possesses significant advantages as a probe for the early properties of the QGP. They can be produced during the initial hard parton scatterings and experience the entire QGP evolution. Consequently, they encode profound information about various medium effects. Furthermore, their large mass scale enables reliable theoretical calculations \cite{demouraQuarkoniumSpinAlignment2023,zhaoHeavyFlavorsExtreme2020,satzColourDeconfinementQuarkonium2006,rothkopfHeavyQuarkoniumExtreme2020}. Therefore, studying the $J/\psi$ spin alignment may advances our understanding of the QGP properties and the interactions of heavy quarks within the medium. Previous measurements in proton-proton ($pp$) collisions yielded no significant $J/\psi$ spin alignment \cite{alicecollaborationMeasurementInclusivepsi2018,andronicHeavyflavourQuarkoniumProduction2016a,adamMeasurementInclusivePolarization2020a}. In contrast, the ALICE Collaboration observed transverse momentum ($p_T$) dependent $J/\psi$ spin alignment in Pb-Pb collisions at the $\sqrt{s_{NN}}=5.02~\mathrm{TeV}$ \cite{acharyaMeasurementPolarizationRespect2023,thealicecollaborationFirstMeasurementVector2025}. The experimental data need theoretical explanation.

In this work, the theoretical framework for deriving the spin density matrix is based on the Kadanoff-Baym (KB) equations within the closed time-path (CTP) formalism \cite{chouEquilibriumNonequilibriumFormalisms1985, cassingKadanoffbaymDynamicsOffshell2009}. This established approach is widely used for finite-temperature non-equilibrium quantum systems. Following the specific methodology developed in Ref.~\cite{shengRelativisticSpinDynamics2024}, we utilize the spin Boltzmann equation for vector mesons. In this formalism, the corresponding phase-space distribution function carries spin indices and is known as the matrix-valued spin dependent distribution (MVSD) \cite{shengKadanoffbaymBoltzmannEquations2021}. As a phase-space distribution, the MVSD serves as the direct counterpart of the observable spin density matrix. While the spin density matrix elements obtained from the spin Boltzmann equation take a relativistic form, adopting a non-relativistic approximation is both natural and essential given our focus on the heavy-flavor meson $J/\psi$. This approximate procedure provides a necessary groundwork for the numerical evaluation of $\rho_{00}$ and further comparisons with experimental results.

The paper is organized as follows. 
In Sec.~\ref{sec2}, the specific form of the spin Boltzmann equation and its formal solution are presented. 
In Sec.~\ref{sec3}, this solution is connected to the spin density matrix elements, and the expressions in both relativistic and non-relativistic cases are given, with particular emphasis on the specific form after integration in the non-relativistic limit.
In Sec.~\ref{sec4}, using the derived non-relativistic formulas, we perform numerical calculations under various conditions to obtain the transverse momentum dependence of $J/\psi$ $\rho_{00}$, and compare our results with experimental data. 
Finally, a summary and discussion of the main results are presented in Sec.~\ref{sec5}.

\textit{Notation.} We use natural units with $\hbar=c=k_B=1$ and the metric convention $g_{\mu \nu}=g^{\mu \nu}=\text{diag}(1, -1, -1, -1)$.

%The spin alignment of light vector mesons $(K^{*0}, \phi)$ was observed first \cite{starcollaborationPatternGlobalSpin2023}, while that of heavy quarkonium (charm quark) was observed later. A relatively large spin alignment signal of $J/\psi$ with respect to the event plane has been measured in Pb-Pb collisions by the ALICE Collaboration \cite{acharyaMeasurementPolarizationRespect2023} as opposed to the previous measurements in pp 
%collisions, where no sizable spin alignment signals for $J/\psi$ were observed at both the LHC \cite{alicecollaborationMeasurementInclusivepsi2018} and RHIC \cite{PhysRevD.102.092009,yangQuarkoniumPolarizationMedium2024}. 

\section{Spin Boltzmann Equation\label{sec2}}
The spin Boltzmann equation for vector mesons reads \cite{shengRelativisticSpinDynamics2024}
\begin{equation}
\frac{p}{E_{p}^{V}}\cdot\partial_{x}f_{\lambda_{1}\lambda_{2}}(x,\mathbf{p})\approx R_{\lambda_{1}\lambda_{2}}^{\mathrm{coal}}(\mathbf{p})-R^{\mathrm{diss}}(\mathbf{p})f_{\lambda_{1}\lambda_{2}}(x,\mathbf{p}),\label{boltzmann}
\end{equation}
where $R_{\lambda_1\lambda_2}^\mathrm{coal}$, $R^\mathrm{diss}$ and $f_{\lambda_1\lambda_2}(x,\mathbf{p})$ denote the coalescence rate, dissociation rates and MVSD for the vector meson, respectively. The coalescence and dissociation rates are defined as \cite{shengRelativisticSpinDynamics2024}
\begin{widetext}
\begin{equation}
\begin{aligned}
R_{\lambda_{1}\lambda_{2}}^{\mathrm{coal}}(\mathbf{p})&=\frac{1}{8(2\pi\ignore{\hbar})^{2}}\sum_{r_{1},s_{1},r_{2},s_{2}}\int d^{3}\mathbf{p}^{\prime}\frac{1}{E_{p^{\prime}}^{\bar{q}}E_{\mathbf{p}-\mathbf{p}^{\prime}}^{q}E_{p}^{V}}\delta(E_{p}^{V}-E_{p^{\prime}}^{\bar{q}}-E_{\mathbf{p}-\mathbf{p}^{\prime}}^{q})\epsilon_{\alpha}^{*}(\lambda_{1},\mathbf{p})\epsilon_{\beta}(\lambda_{2},\mathbf{p})\\&\times\mathrm{Tr}\left[\Gamma^{\beta}v(s_{1},\mathbf{p}^{\prime})\bar{v}(r_{1},\mathbf{p}^{\prime})\Gamma^{\alpha}u(r_{2},\mathbf{p}-\mathbf{p}^{\prime})\bar{u}(s_{2},\mathbf{p}-\mathbf{p}^{\prime})\right]f_{r_{1}s_{1}}^{(-)}(x,\mathbf{p}^{\prime})f_{r_{2}s_{2}}^{(+)}(x,\mathbf{p}-\mathbf{p}^{\prime}),\label{Rcoal}
\end{aligned}
\end{equation}
\begin{equation}
\begin{aligned}
R^{\mathrm{diss}}(\mathbf{p})&=-\frac{1}{12(2\pi\ignore{\hbar})^{2}}\sum_{r_{1},r_{2}}\int d^{3}\mathbf{p}^{\prime}\frac{1}{E_{p^{\prime}}^{\bar{q}}E_{\mathbf{p}-\mathbf{p}^{\prime}}^{q}E_{p}^{V}}\delta(E_{p}^{V}-E_{p^{\prime}}^{\bar{q}}-E_{\mathbf{p}-\mathbf{p}^{\prime}}^{q})\left(g_{\alpha\beta}-\frac{p_{\alpha}p_{\beta}}{m_{V}^{2}}\right)\\&\times\mathrm{Tr}\left\{\Gamma^{\beta}(p^{\prime}\cdot\gamma-m_{\bar{q}})\Gamma^{\alpha}\left[(p-p^{\prime})\cdot\gamma+m_{q}\right]\right\},\label{Rdiss}
\end{aligned}
\end{equation}
\end{widetext}
where $\Gamma^\alpha$ denotes the $q\bar{q}V$ vertex:
\begin{equation}
    \Gamma^\alpha\approx g_VB(\mathbf{p}-\mathbf{p}^{\prime},\mathbf{p}^{\prime})\gamma^\alpha
\end{equation}
and $B(\mathbf{p}-\mathbf{p}^{\prime},\mathbf{p}^{\prime})$ represents the Bethe-Salpeter wave function of the meson \cite{xuVectormesonProductionVector2021}. The coupling constant $g_V$ characterizes the strength of interaction between vector meson and quark-antiquark. In the definition of coalescence rate (Eq.~\ref{Rcoal}), $u/v$ and $\bar{u}/\bar{v}$ are Dirac spinors for quarks/antiquarks, and $f_{rs}^{(\pm)}$ are the MVSD of quarks and antiquarks \cite{shengRelativisticSpinDynamics2024,shengSpinAlignmentVector2023}, which reads
\begin{equation}
    f_{rs}^{q/\bar{q}}(x,\mathbf{p})=\frac{1}{2}f_{q/\bar{q}}(x,\mathbf{p})[\delta_{rs}-P_{\mu}^{q/\bar{q}}(x,\mathbf{p})n_{j}^{\mu}(\mathbf{p})\tau_{rs}^{j}],
\end{equation}
where $P_{\mu}^{q/\bar{q}}(x,\mathbf{p})$ are polarization four-vectors of (anti-)quarks, $n_{j}^{\mu}(\mathbf{p})$ ($j = 1,2,3$) are four-vectors of three basis directions for spin states in the rest frame of (anti-)quarks, and $\tau^{j}$ ($j = 1,2,3$) are Pauli matrices. Moreover, $\epsilon_\mu(\lambda,\mathbf{p})$ represents the polarization four-vector of vector mesons.

The formal solution to Eq.~\ref{boltzmann} reads \cite{shengRelativisticSpinDynamics2024}
\begin{equation}
\begin{aligned}
f_{\lambda_{1}\lambda_{2}}(x,\mathbf{p})&\sim\frac{R_{\lambda_{1}\lambda_{2}}^{\mathrm{coal}}(\mathbf{p})}{R^{\mathrm{diss}}(\mathbf{p})}\left[1-\exp{(-R^{\mathrm{diss}}(\mathbf{p})\Delta t)}\right]
%\\&\left.\sim\left\{\begin{array}{ll}{R_{\lambda_{1}\lambda_{2}}^{\mathrm{coal}}(\mathbf{p})\Delta t,}&{\mathrm{for}\Delta t\ll1/R^{\mathrm{diss}}(\mathbf{p})}\\\\\frac{R_{\lambda_{1}\lambda_{2}}^{\mathrm{coal}}(\mathbf{p})}{R^{\mathrm{diss}}(\mathbf{p})}&{\mathrm{for}\Delta t\gg1/R^{\mathrm{diss}}(\mathbf{p})}\end{array}\right.\right.
\end{aligned}
\label{f12}
\end{equation}
when $f_{\lambda_{1}\lambda_{2}}(x,\mathbf{p})$ at the initial time is assumed to be zero, where $\Delta t$ is the formation time of the vector meson. If the coalescence process of the vector meson is assumed to take place within a very short time period, i.e., $\Delta t \ll 1/R^{\mathrm{diss}}(\mathbf{p})$, then the following approximate result is obtained
\begin{equation}
    f_{\lambda_{1}\lambda_{2}}(x,\mathbf{p})\sim R_{\lambda_{1}\lambda_{2}}^{\mathrm{coal}}(\mathbf{p})\Delta t.\label{fapprox}
\end{equation}
This result is connected with the spin density matrix to be discussed.

\section{Spin Density Matrix For Mesons\label{sec3}}
\subsection{Relativistic case}
Our ultimate goal is to derive the expression for the spin density matrix of $J/\psi$ particles in the non-relativistic limit. As the approach adopted in this paper is based on relativistic theory, some results in the relativistic case are essential.

The formal solution of the vector particle MVSD in the spin Boltzmann equation is given in Eq.~\ref{fapprox}. It reveals the dynamical evolution of the vector meson's spin states within the medium. The coalescence rate $R^{\mathrm{coal}}_{\lambda_1\lambda_2}(\mathbf{p})$ intrinsically possesses a matrix structure in the spin space. Therefore $f_{\lambda_1\lambda_2}(x, \mathbf{p})$ naturally constitutes a Hermitian matrix that fully describes the statistical properties of the vector meson's mixed states. This effectively allows $f_{\lambda_1\lambda_2}(x, \mathbf{p})$ to act as the unnormalized spin density matrix in phase space \cite{shengSpinAlignmentVector2023}. Thus the unnormalized spin density matrix can be represented as the following compact form \cite{shengRelativisticSpinDynamics2024}
\begin{widetext}
\begin{equation}
\begin{aligned}
\rho_{\lambda_{1}\lambda_{2}}^{V}(x,\mathbf{p})&=\frac{\Delta t}{32}\int\frac{d^{3}\mathbf{p}^{\prime}}{(2\pi\ignore{\hbar})^{3}}\frac{1}{E_{p^{\prime}}^{\bar{q}}E_{\mathbf{p}-\mathbf{p}^{\prime}}^{q}E_{p}^{V}}f_{\bar{q}}(x,\mathbf{p}^{\prime})f_{q}(x,\mathbf{p}-\mathbf{p}^{\prime})2\pi\ignore{\hbar}\delta(E_{p}^{V}-E_{p^{\prime}}^{\bar{q}}-E_{\mathbf{p}-\mathbf{p}^{\prime}}^{q})\epsilon_{\alpha}^{*}(\lambda_{1},\mathbf{p})\epsilon_{\beta}(\lambda_{2},\mathbf{p})\\&\times\mathrm{Tr}\left\{\Gamma^{\beta}(p^{\prime}\cdot\gamma-m_{\tilde{q}})\left[1+\gamma_{5}\gamma\cdot P^{\tilde{q}}(x,\mathbf{p}^{\prime})\right]\Gamma^{\alpha}\left[(p-p^{\prime})\cdot\gamma+m_{q}\right]\left[1+\gamma_{5}\gamma\cdot P^{q}(x,\mathbf{p}-\mathbf{p}^{\prime})\right]\right\},\end{aligned}   
\end{equation}
\end{widetext}
in which $P^{q/\bar{q}}(x,\mathbf{p})$ are \cite{shengWhatCanWe2020,shengImprovedQuarkCoalescence2020,shengSpinAlignmentVector2023,fangPolarizationMassiveFermions2016,yangQuarkCoalescenceModel2018}
\begin{equation}
\begin{aligned}
    P_{q}^{\mu}(x,\mathbf{p})=&\frac{1}{2m_q}\left(\tilde{\omega}^{\mu\nu}+\frac{g_{V}}{E_{p}^qT}\tilde{F}_{V}^{\mu\nu}\right)p_{\nu}\\&\times[1-f_{q}(x,\mathbf{p})],\\
    P_{\bar{q}}^{\mu}(x,\mathbf{p})=&\frac{1}{2m_{\bar{q}}}\left(\tilde{\omega}^{\mu\nu}-\frac{g_{V}}{E_{p}^{\bar{q}}T}\tilde{F}_{V}^{\mu\nu}\right)p_{\nu}\\&\times[1-f_{\bar{q}}(x,\mathbf{p})].
\end{aligned}
\label{po4v}
\end{equation}
where $\tilde{\omega}^{\mu\nu}$  is the dual of the thermal vorticity tensor \cite{Huang2021}, and $\tilde{F}_{V}^{\mu\nu}$ is the dual of the
field strength tensor of vector mesons.  In addition, $p_\nu=(E_p^{q/\bar{q}},\mathbf{p})$ denote the on-shell momentum of (anti-)quarks with $E_p^{q/\bar{q}}=\sqrt{\mathbf{p^2}+m_{q/\bar{q}}^2}$. The energy of (anti-)quarks is obtained from the contraction of the local fluid velocity and momentum, evaluated in the rest frame of the vector meson where $u^\mu = (1,0,0,0)$ ($u^\mu$ is the local fluid velocity). 
Here in polarization four-vectors $f_{q/\bar{q}}(x,\mathbf{p})$ are the Fermi-Dirac distributions of (anit-)quarks, which reads
\begin{equation}
    f_{q/\bar{q}}(x,\mathbf{p})=\frac{1}{1+\exp(\frac{E_{p}^{q/\bar{q}}\mp\mu_{q}}{T})},
\end{equation}
where $T$ is the coalescence temperature and $\mu_q$ is the chemical potential. In the scenario where we will perform our calculations, the value of the Fermi-Dirac distribution $f_{FD}$ is much less than 1 and can therefore be neglected \cite{shengSpinAlignmentVector2023}.

In the relativistic case, according to the aforementioned polarization four-vector, the spin density matrix of the vector meson may contain the following momentum integral structure \cite{shengRelativisticSpinDynamics2024}:
\begin{equation}
    \begin{aligned}
\left\{I,I^{\mu},I^{\mu\nu},I^{\mu\nu\rho},I^{\mu\nu\rho\sigma}\right\}&=\int\frac{d^{3}\mathbf{p}^{\prime}}{(2\pi\ignore{\hbar})^{3}}\frac{1}{E_{p^{\prime}}^{\bar{q}}E_{\mathbf{p}-\mathbf{p}^{\prime}}^{q}}\\&\times B^{2}(\mathbf{p}-\mathbf{p}^{\prime},\mathbf{p}^{\prime})f_{\bar{q}}(\mathbf{p}^{\prime})f_{q}(\mathbf{p}-\mathbf{p}^{\prime})\\&\times2\pi\ignore{\hbar}\delta(E_{p}^{V}-E_{p^{\prime}}^{\bar{q}}-E_{\mathbf{p}-\mathbf{p}^{\prime}}^{q})\\&\times\left\{1,p^{\prime\mu},p^{\prime\mu}p^{\prime\nu},p^{\prime\mu}p^{\prime\nu}p^{\prime\rho},p^{\prime\mu}p^{\prime\nu}p^{\prime\rho}p^{\prime\sigma}\right\}.
    \end{aligned}
\label{integral}
\end{equation}
Now, the fundamental results in the relativistic case have been obtained. Next, we will derive the normalized spin density matrix in the non-relativistic case on this basis, i.e., 
\begin{equation}
    \bar{\rho}_{\lambda_1\lambda_2}^V(x,\mathbf{p})=\frac{\rho_{\lambda_1\lambda_2}^V(x,\mathbf{p})}{\mathrm{Tr}(\rho_V)}.
\end{equation}

\subsection{Non-relativistic limit}
For heavy flavor particles, non-relativistic limit is a good approximation. In such limit, one consider $m_{q}=m_{\bar{q}}$ and assume $m_{V}\approx 2m_{q}$. Therefore the following approximation can be taken
\begin{equation}
\begin{aligned}
p^{\mu}&\approx(m_{V},\mathbf{0}),\\
p^{\prime\mu}&\approx(m_{\bar{q}},\mathbf{0})=(m_{q},\mathbf{0}),\\
p^{\mu}-p^{\prime\mu}&\approx(m_{q},\mathbf{0}).
\end{aligned}
\end{equation}
Since the polarization four-vector and momentum are required to be orthogonal, the non-relativistic approximation of the polarization four-vectors for quarks and vector mesons can be obtained accordingly:
\begin{equation}
    \begin{gathered}P_{\bar{q}}^{\mu}(x,\mathbf{p}^{\prime})\approx(0,\mathbf{P}_{\bar{q}}(x,\mathbf{p}^{\prime})),\\P_{q}^{\mu}(x,\mathbf{p}-\mathbf{p}^{\prime})\approx(0,\mathbf{P}_{q}(x,\mathbf{p}-\mathbf{p}^{\prime})),\\\epsilon^{\mu}(\lambda_{1})\approx(0,\epsilon(\lambda_{1})),\\\epsilon^{\mu}(\lambda_{2})\approx(0,\epsilon(\lambda_{2})).\end{gathered}
\end{equation}

The polarization three-vector $\epsilon(\lambda)$ are \cite{shengRelativisticSpinDynamics2024}
\begin{equation}
    \begin{aligned}
    \epsilon(0)&=(0,1,0),\\
    \epsilon(+1)&=-\frac{1}{\sqrt{2}}(i,0,1),\\
    \epsilon(-1)&=\frac{1}{\sqrt{2}}(-i,0,1).
    \end{aligned}
\end{equation}

\begin{widetext}
Then the spin density matrix for the vector meson in the non-relativistic limit is given by
\begin{equation}
\begin{aligned}
    \rho_{\lambda_{1}\lambda_{2}}^{V}(x,\mathbf{p})&=\frac{\Delta t}{8}g_{V}^{2}m_{V}m_{q}\int\frac{d^{3}\mathbf{p}^{\prime}}{(2\pi\ignore{\hbar})^{3}}\frac{1}{E_{p^{\prime}}^{\bar{q}}E_{\mathbf{p}-\mathbf{p}^{\prime}}^{q}E_{p}^{V}}f_{\bar{q}}(x,\mathbf{p}^{\prime})f_{q}(x,\mathbf{p}-\mathbf{p}^{\prime})2\pi\ignore{\hbar}\delta(E_{p}^{V}-E_{p^{\prime}}^{\bar{q}}-E_{\mathbf{p}-\mathbf{p}^{\prime}}^{q})\\&
    \times\Big\{\boldsymbol{\epsilon}^*(\lambda_1)\cdot\boldsymbol{\epsilon}(\lambda_2)\left[1+\mathbf{P}_q(x,\mathbf{p}-\mathbf{p}^{\prime})\cdot\mathbf{P}_{\bar{q}}(x,\mathbf{p}^{\prime})\right]-\left[\mathbf{P}_q(x,\mathbf{p}-\mathbf{p}^{\prime})\cdot\boldsymbol{\epsilon}(\lambda_2)\right]\left[\mathbf{P}_{\bar{q}}(x,\mathbf{p}^{\prime})\cdot\boldsymbol{\epsilon}^*(\lambda_1)\right]\\&
    -\left[\mathbf{P}_q(x,\mathbf{p}-\mathbf{p}^{\prime})\cdot\boldsymbol{\epsilon}^*(\lambda_1)\right]\left[\mathbf{P}_{\bar{q}}(x,\mathbf{p}^{\prime})\cdot\boldsymbol{\epsilon}(\lambda_2)\right]-i\left[\boldsymbol{\epsilon}^*(\lambda_1)\times\boldsymbol{\epsilon}(\lambda_2)\right]\cdot\left[\mathbf{P}_q(x,\mathbf{p}-\mathbf{p}^{\prime})+\mathbf{P}_{\bar{q}}(x,\mathbf{p}^{\prime})\right]\Big\}.
\end{aligned}
\label{rho00nonrel}
\end{equation}
\end{widetext}
The above formula can be simplified by using the shorthand notation
\begin{equation}
    \begin{aligned}
    Dp^{\prime}&\equiv\frac{\Delta t}{8}g_{V}^{2}m_{V}m_{q}\int\frac{d^{3}\mathbf{p}^{\prime}}{(2\pi\ignore{\hbar})^{3}}\frac{1}{E_{p^{\prime}}^{\bar{q}}E_{\mathbf{p-p^{\prime}}}^{q}E_{p}^{V}}\\&\times f_{\bar{q}}(x,\mathbf{p}^{\prime})f_{q}(x,\mathbf{p}-\mathbf{p}^{\prime})2\pi\ignore{\hbar}\delta(E_{p}^{V}-E_{p^{\prime}}^{\bar{q}}-E_{\mathbf{p-p^{\prime}}}^{q}).\end{aligned}
    \end{equation}
The 00-element of the normalized density matrix is given by
\begin{equation}
\begin{aligned}
\bar{\rho}_{00}&=\frac{\rho_{00}}{\rho_{11}+\rho_{00}+\rho_{-1,-1}}\\&=\frac{\int Dp^{\prime}\left[1+\mathbf{P}_{q}\cdot\mathbf{P}_{\bar{q}}-2(\mathbf{n}_{3}\cdot\mathbf{P}_{q})(\mathbf{n}_{3}\cdot\mathbf{P}_{\bar{q}})\right]}{\int Dp^{\prime}(3+\mathbf{P}_{q}\cdot\mathbf{P}_{\bar{q}})},\end{aligned}
\end{equation}
where $\mathbf{P}_q=\mathbf{P}_q(x,\mathbf{p-p^\prime})$ and $\mathbf{P}_{\bar{q}}=\mathbf{P}_{\bar{q}}(x,\mathbf{p^\prime})$.
If the polarization value is very small compared to unity, then the Taylor expansion can be applied to the above density matrix element,
\begin{equation}
\begin{aligned}
\bar{\rho}_{00}(x,\mathbf{p})\simeq&\frac{1}{3}+\frac{2}{9N}\int Dp^{\prime}\big[(\mathbf{n}_{1}\cdot\mathbf{P}_{q})(\mathbf{n}_{1}\cdot\mathbf{P}_{\bar{q}})\\&+(\mathbf{n}_{2}\cdot\mathbf{P}_{q})(\mathbf{n}_{2}\cdot\mathbf{P}_{\bar{q}})-2(\mathbf{n}_{3}\cdot\mathbf{P}_{q})(\mathbf{n}_{3}\cdot\mathbf{P}_{\bar{q}})\big],
\end{aligned}
\end{equation}
where $N\equiv\int Dp'$ is the normalization constant, $\mathbf{n}_1$, $\mathbf{n}_2$ and $\mathbf{n}_3$ are orthogonal basis vectors in the rest frame of the vector meson \cite{shengRelativisticSpinDynamics2024}.
Under the simplifying assumption that $\mathbf{P}_q$ and $\mathbf{P}_{\bar{q}}$ have nonzero components only along $\mathbf{n}_3$, the expression above becomes
\begin{equation}
\begin{aligned}\bar{\rho}_{00}(x,\mathbf{p})\approx\frac{1}{3}-\frac{4}{9N}\int Dp^{\prime}&\left[\mathbf{n}_{3}\cdot\mathbf{P}_{q}(x,\mathbf{p}-\mathbf{p}^{\prime})\right]\\&\times\left[\mathbf{n}_{3}\cdot\mathbf{P}_{\bar{q}}(x,\mathbf{p}^{\prime})\right].
\end{aligned}
\label{normrho00}
\end{equation}
To get the complete form of this density matrix element, we need the quark polarization vector \cite{shengImprovedQuarkCoalescence2020,shengWhatCanWe2020,fangPolarizationMassiveFermions2016},
\begin{equation}
    \begin{aligned}
    \mathbf{P}_{\mathrm{q/\bar{q}}}^{y}(x,\mathbf{p})&=\frac{1}{2}\mathbf{\omega}_{y}\pm\frac{1}{2m_{\mathrm{q}}}(\mathbf{\varepsilon}\times\mathbf{p})_{y}\\
    &\pm\frac{g_{V}}{2m_{\mathrm{q}}T}\mathbf{B}_{y}^{V}+\frac{g_{V}}{2m_{\mathrm{q}}E_{p}^{q/\bar{q}}T}(\mathbf{E}^{V}\times\mathbf{p})_{y}.
    \label{quarkpolarization}
    \end{aligned}
\end{equation}
when taking $\mathbf{n}_{3}=(0,1,0)$ (along the $y$ axis) as the spin quantization direction. Here $\mathbf{\omega}$ and $\mathbf{\varepsilon}$ are the magnetic part and electric part of the thermal vorticity tensor. Similarly, $\mathbf{B}^V$ and $\mathbf{E}^V$ are the magnetic part and electric part of the vector field. Equations \ref{rho00nonrel}-\ref{normrho00} above correspond, in order, to Eqs. (C4), (C5), (C9), (C10), and (C12) in Appendix C of Ref. \cite{shengRelativisticSpinDynamics2024}.

It should be noted that the vector field represented by the field strength tensor here is not the same as the vector meson field in Eq.~\ref{po4v}. We assume that the vector field responsible for the polarization of heavy quarks is the gluon field and its fluctuations, rather than the vector field induced by the pseudo-Goldstone boson current \cite{shengRelativisticSpinDynamics2024}. As a phenomenological ansatz, we treat it as an effective Abelian-like vector field here, with the field strength tensor taking the same form as that of the vector meson field. This treatment allows us to consistently implement the relativistic spin Boltzmann framework. Consequently, the coupling constant $g_V$ no longer represents the quark-meson coupling, but rather characterizes the effective interaction strength between the heavy quarks and the local gluon field fluctuations. 

After taking an averaging over spacetime, one can obtain
\begin{widetext}
    \begin{equation}
    \begin{aligned}
     \langle\bar{\rho}_{00}(x,\mathbf{p})\rangle_x\approx\frac{1}{3}&-\frac{4}{9N}\int Dp^\prime \Big\{\frac{1}{4}\langle\mathbf{\omega}_y^2\rangle-\frac{1}{4m_q^2}\langle\big[\mathbf{\varepsilon}\times(\mathbf{p}-\mathbf{p^\prime})\big]_y\big[\mathbf{\varepsilon}\times(\mathbf{p^\prime})\big]_y\rangle\\&
    -\frac{g_V^2}{4m_q^2T^2}\langle\mathbf{B}_{Vy}^2\rangle+\frac{g_V^2}{4m_q^2T^2E_p^{V2}}\langle\big[\mathbf{E}_V\times(\mathbf{p}-\mathbf{p^\prime})\big]_y\big[\mathbf{E}_V\times(\mathbf{p^\prime})\big]_y\rangle\Big\}.
    \end{aligned}
\end{equation}
\end{widetext}
It's convenient to calculate in the vector meson rest frame, which means $\mathbf{p}=\mathbf{0}$. Note that there is an integral structure in the above equation similar to that in Eq.~\ref{integral}:
\begin{equation}
    \begin{aligned}
\left\{I,I^{\mu\nu}\right\}&=\int\frac{d^{3}\mathbf{p}^{\prime}}{(2\pi\ignore{\hbar})^{3}}\frac{1}{E_{p^{\prime}}^{\bar{q}}E_{\mathbf{p}-\mathbf{p}^{\prime}}^{q}}\\&\times B^{2}(\mathbf{p}-\mathbf{p}^{\prime},\mathbf{p}^{\prime})f_{\bar{q}}(\mathbf{p}^{\prime})f_{q}(\mathbf{p}-\mathbf{p}^{\prime})\\&\times2\pi\ignore{\hbar}\delta(E_{p}^{V}-E_{p^{\prime}}^{\bar{q}}-E_{\mathbf{p}-\mathbf{p}^{\prime}}^{q})\left\{1,p^{\prime\mu}p^{\prime\nu}\right\},
    \end{aligned}
\end{equation}
which spatial component is $I^{ij}=(1/3)\delta_{ij}(m_V^2/4-m_q^2)I$.
Meanwhile, remember that in non-relativistic limit $m_V\approx2m_q$, after simplification the spacetime average of this 00-element becomes
\begin{equation}
    \langle\bar{\rho}_{00}(x,\mathbf{p})\rangle_x\approx\frac{1}{3}-\frac{1}{9}\big[\langle\mathbf{\omega}_y^{\prime2}\rangle-\frac{g_V^2}{m_q^2T^2}\langle\mathbf{B}_{Vy}^{\prime2}\rangle\big].
    \label{rho00origin}
\end{equation}

This formula was derived in the rest frame of the vector particle. To restore its momentum dependence, we now switch to the lab frame for the calculation. The
Lorentz transformation is as follows \cite{shengRelativisticSpinDynamics2024,shengMomentumDependenceSpin2023}:
\begin{equation}
\begin{aligned}
     \mathbf{\omega^{\prime}}&=\gamma\mathbf{\omega}-\gamma\mathbf{v}\times\mathbf{\varepsilon}+(1-\gamma)\frac{\mathbf{v}\cdot\boldsymbol{\omega}}{v^{2}}\mathbf{v},\\
     \mathbf{B}_{V}^{\prime}&=\gamma\mathbf{B}_{V}-\gamma\mathbf{v}\times\mathbf{E}_{V}+(1-\gamma)\frac{\mathbf{v}\cdot\mathbf{B}_{V}}{v^{2}}\mathbf{v},
\end{aligned}
\end{equation}
where the primed quantities refer to the rest frame and the unprimed ones to the laboratory frame and $\gamma$ the Lorentz factor. To obtain the $p_T$ dependence of $J/\psi$’s $\rho_{00}^V$, one needs to re-express the momentum components and the energy in terms of the transverse momentum $p_T$, the polar angle $\phi$, and the rapidity $Y$. The explicit form is as follows:
\begin{equation}
    \begin{aligned}
        p_x&=p_T~\mathrm{cos}(\phi),\\
        p_y&=p_T~\mathrm{sin}(\phi),\\
        p_z&=\sqrt{m_V^2+p_T^2}~\mathrm{sinh}(Y),\\
        E_\mathbf{p}^V&=\sqrt{m_V^2+p_T^2}~\mathrm{cosh}(Y).
    \end{aligned}
\end{equation}
After the change of variables, we perform an average of $\bar{\rho}_{00}$ over the polar angle and rapidity, weighted by the momentum spectrum of the $J/\psi$ mesons,
\begin{equation}
    E_{\mathbf{p}}^V\frac{d^3N}{d^3\mathbf{p}}=\frac{d^2N}{2\pi p_Tdp_Tdy}[1+2v_2(p_T)\cos(2\phi)],
\end{equation}
where the elliptic flow $v_2(p_T)$ is taken from ALICE's data \cite{thealicecollaborationEllipticTriangularFlow2020}.

\begin{widetext}
Finally, one can obtain
    \begin{equation}
    \begin{aligned}
    \langle{\bar{\rho}}_{00}(p_T)\rangle_{x,\phi,Y}&\approx\langle\frac{1}{3}-\frac{1}{9}\Big\{\langle\omega_y^2\rangle\Big[(F^2+\frac{1}{2}G^2)+2FG~v_2(p_T)\Big]\\&-f_T^2\Big[(F^2+\frac{1}{2}G^2)+2FG~v_2(p_T)+\frac{\left(m_V^{2}+p_T^{2}\right)}{m_V^{2}}\mathrm{sinh}(Y)^{2}+\frac{G^2}{2}\Big]\\&
    -f_z^2\Big[\frac{p_T^{2}}{m_V^{2}}\left(\frac{1}{2}+\frac{1}{2}v2(p_T)\right)
    +\frac{p_T^{2}\left(m_V^{2}+p_T^{2}\right)}{m_V^{2}\left(\sqrt{m_V^{2}+p_T^{2}}~\mathrm{cosh}(Y)+m_V\right)^{2}}\left(\frac{1}{2}-\frac{1}{2}v_2(p_T)\right)~\mathrm{sinh}(Y)^{2}\Big]\Big\}\rangle_Y,
    \end{aligned}\label{rho00boostnorm}
\end{equation}
\end{widetext}

with the definition,
\begin{align}
        G&=\frac{p_T^2}{2m_V(\sqrt{m_V^2+p_T^2}~\mathrm{cosh}(Y)+m_V)},
        \end{align}
        \begin{align}
        F&=-G+\frac{\sqrt{m_V^2+p_T^2}~\mathrm{cosh}(Y)}{m_V}.
\end{align}

Here in the evaluation, the contributions from the electric part of thermal vorticity $\boldsymbol{\varepsilon}$ are neglected, as the hydrodynamic simulations indicate that its numerical magnitude is negligibly small during the QGP evolution \cite{pangPseudorapidityDistributionDecorrelation2018,pangEffectsInitialFlow2012,shengWhatCanWe2020,shengMomentumDependenceSpin2023}. Meanwhile, for the magnetic part of thermal vorticity $\boldsymbol{\omega}$, the components perpendicular to the huge orbital angular momentum of QGP ($\boldsymbol{\omega}_x$ and $\boldsymbol{\omega}_z$) are also naturally suppressed. Therefore, the spacetime averages of both $\boldsymbol{\varepsilon}_i^2$ and $\boldsymbol{\omega}_{x,z}^2$  can be justifiably omitted from the dominant contributions.

For the vector field in Eq.~\ref{rho00boostnorm}, the correlation between its magnetic and electric parts is neglected, i,e, $\langle B^V_i E_j^V \rangle=0$ \cite{shengMomentumDependenceSpin2023}. Considering the geometry of the QGP, the transverse and longitudinal fields are assumed to be different here, and therefore their fluctuations are characterized by distinct parameters: $f_T^2=\langle\frac{g_V^2}{m_q^2T^2}\mathbf{B}_{Vx,y}^2\rangle=\langle\frac{g_V^2}{m_q^2T^2}\mathbf{E}_{Vx,y}^2\rangle$ and $f_z^2=\langle\frac{g_V^2}{m_q^2T^2}\mathbf{B}_{Vz}^2\rangle=\langle\frac{g_V^2}{m_q^2T^2}\mathbf{E}_{Vz}^2\rangle$ \cite{shengMomentumDependenceSpin2023,shengSpinAlignmentVector2023}.

\section{Numerical results\label{sec4}}

\begin{figure}
    \centering
    \includegraphics[width=1\linewidth]{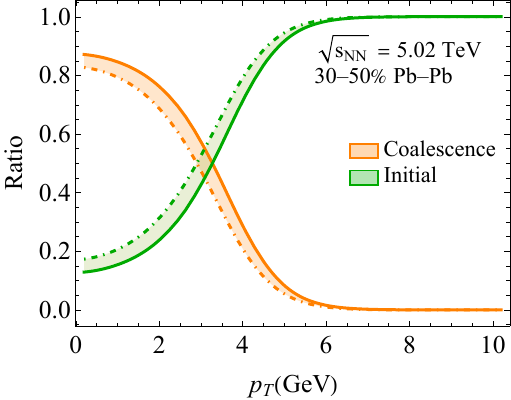}
    \caption{
    Model-calculated relative fractions of the coalescence and initial-production components in the prompt $J/\psi$ yield for Pb-Pb collisions at $\sqrt{s_{\mathrm{NN}}}=5.02$ TeV and 30--50\% centrality. The dash-dotted lines correspond to a nuclear shadowing factor of $0.60$, while the solid lines correspond to $0.85$.
}
    \label{ratiofig}
\end{figure}

In this section, we present numerical results for the $p_T$ dependence of $\rho_{00}$ for $J/\psi$ mesons under different production mechanisms. The available experimental data from the LHC correspond to inclusive measurements at forward rapidity in heavy-ion collisions. In this kinematic region, the final observed $J/\psi$ yield is expected to originate from a mixture of different sources. Therefore, to describe the experimental data within the present phenomenological framework, we introduce a $p_T$-dependent superposition of the coalescence and initial-production components. Based on this, the final observable $\rho_{00}$ can be simplified as
\begin{equation}
    \rho_{00}^{\mathrm{obs}}(p_T)
    =
    r^{\mathrm{coal}}(p_T)\rho_{00}^{\mathrm{coal}}(p_T)
    +
    r^{\mathrm{init}}(p_T)\rho_{00}^{\mathrm{init}},
    \label{combination}
\end{equation}
where $r^{\mathrm{coal}}(p_T)$ and $r^{\mathrm{init}}(p_T)$ denote the relative yield fractions of the coalescence and initial-production processes, respectively, satisfying the normalization condition
\begin{equation}
    r^{\mathrm{coal}}(p_T)+r^{\mathrm{init}}(p_T)=1 .
\end{equation}
Here the two components are assumed to provide the dominant prompt contribution to the measured $J/\psi$ sample in the present phenomenological treatment. The superscripts ``coal'', ``init'', and ``obs'' denote the quantities corresponding to the coalescence process, initial production, and final total observable, respectively. The numerical results for the relative fractions are calculated with the transport model \cite{zhouMediumEffectsCharmonium2014} and presented in Fig.~\ref{ratiofig}, where the uncertainty associated with the nuclear shadowing factor~\cite{k.j.eskolaEPS09NewGeneration2009}. As illustrated in the figure, $J/\psi$ mesons produced via the coalescence mechanism dominate at low transverse momentum, while the fraction from initial production gradually increases and becomes dominant at higher $p_T$.

The coalescence component is assumed to inherit the spin polarization of charm and anticharm quarks generated in the QGP. The initial-production component, on the other hand, is treated as an effective high-$p_T$ baseline. In the present calculation, we fix $\rho_{00}^{\mathrm{init}}=0.36$, motivated by the highest-$p_T$ ALICE data point, where the initial-production fraction is expected to dominate. This value should be regarded as an effective phenomenological input rather than a first-principles prediction of the primordial $J/\psi$ spin alignment.

We first consider the case without the vector-field contribution. In this case, the spin alignment of the coalescence component is induced only by the thermal vorticity of the QGP. From Eq.~(\ref{rho00boostnorm}), by setting $f_T=f_z=0$, one obtains
\begin{equation}
\begin{aligned}
    \rho_{00}^{\mathrm{coal}}(p_T)
    &=
    \langle{\bar{\rho}}_{00}(p_T)\rangle_{x,\phi,Y} \\
    &\approx
    \left\langle
    \frac{1}{3}
    -
    \frac{1}{9}
    \langle\omega_y^2\rangle
    \left[
    \left(F^2+\frac{1}{2}G^2\right)
    +
    2FG\,v_2(p_T)
    \right]
    \right\rangle_Y ,
\end{aligned}
\label{rho00thermal}
\end{equation}
where the subscript $Y$ denotes the rapidity average in the corresponding rapidity window. The thermal vorticity acts on the spin degrees of freedom of the charm and anticharm quarks before they coalesce into a $J/\psi$ meson. Thus, its effect is most visible in the region where the coalescence fraction is sizable. Since the averaged squared thermal vorticity is not determined independently in the present framework, we calibrate it using the lowest-$p_T$ forward-rapidity ALICE data poin. The uncertainty in the nuclear shadowing factor leads to the uncertainty in $\langle\omega_y^2\rangle$.

\begin{figure}[htb]
    \centering
    \includegraphics[width=1\linewidth]{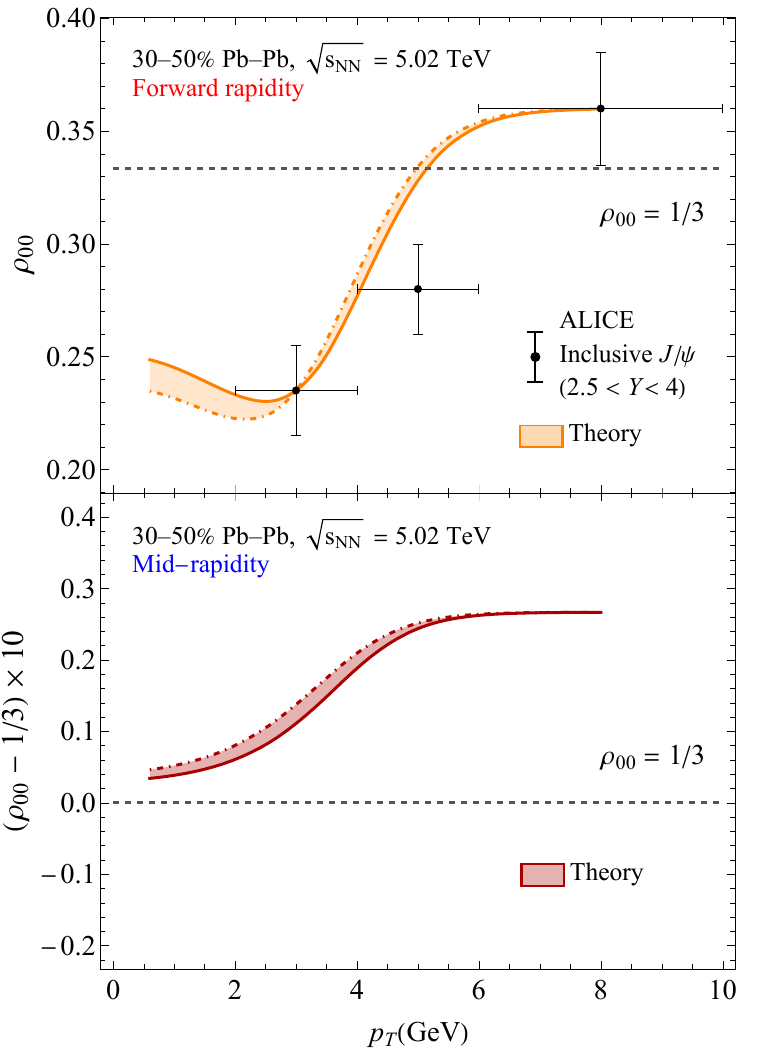}
    \caption{
The $J/\psi$ spin density matrix element as a function of transverse momentum $p_T$ without the vector-field contribution. The upper panel shows $\rho_{00}$ at forward rapidity, while the lower panel shows $(\rho_{00}-1/3)\times 10$ at mid-rapidity. The widths of the bands reflect the uncertainty from the nuclear shadowing factor. The dash-dotted and solid lines correspond to shadowing factors of $0.60$ and $0.85$, respectively. The horizontal grey dashed line indicates the unpolarized reference value $\rho_{00}=1/3$. Experimental data from the ALICE Collaboration \cite{thealicecollaborationFirstMeasurementVector2025} are shown as black points with error bars in the upper panel.
}
    \label{Fullfig}
\end{figure}

The results without the vector-field contribution are shown in Fig.~\ref{Fullfig}. In the forward-rapidity region, the calculated $\rho_{00}$ exhibits a non-monotonic dependence on $p_T$. At low and intermediate $p_T$, the coalescence contribution is important, and the thermal-vorticity-induced charm-quark polarization drives $\rho_{00}^{\mathrm{coal}}$ below the unpolarized value $1/3$. As $p_T$ increases, the initial-production fraction grows rapidly, and the total $\rho_{00}^{\mathrm{obs}}$ is pulled toward the effective initial-production value $\rho_{00}^{\mathrm{init}}=0.36$. Therefore, the decrease and subsequent increase of the forward-rapidity curve should be understood as the result of the competition between the two production components, with additional partially spin polarized charm quarks. 

The lower panel of Fig.~\ref{Fullfig} shows the corresponding result at mid-rapidity. To make the small variation around the unpolarized reference visible, the plotted quantity is $(\rho_{00}-1/3)\times 10$. Compared with the forward-rapidity result, the direct thermal-vorticity-induced deviation is much weaker at mid-rapidity. This suppression is mainly kinematic in origin. The spin density matrix is naturally evaluated in the $J/\psi$ rest frame, whereas the experimental variables $p_T$ and $Y$ are defined in the laboratory frame. The Lorentz transformation in Eq.~(\ref{rho00boostnorm}) makes the vorticity contribution sensitive to the longitudinal momentum of the $J/\psi$. Consequently, the same thermal-vorticity source can produce a visible spin-alignment effect at forward rapidity, while its contribution becomes strongly suppressed near mid-rapidity, where the longitudinal momentum is small.

As a result, the mid-rapidity curve in Fig.~\ref{Fullfig} is controlled mainly by the changing production fractions. At low $p_T$, the coalescence component dominates and its $\rho_{00}^{\mathrm{coal}}$ remains close to $1/3$ because the thermal-vorticity effect is kinematically suppressed. At high $p_T$, the initial-production component becomes dominant and drives the total result toward $\rho_{00}^{\mathrm{init}}=0.36$. Thus, Fig.~\ref{Fullfig} illustrates two intertwined effects: the yield-fraction competition between coalescence and initial production, and the rapidity-dependent kinematic response of the thermal-vorticity contribution.

\begin{figure}[htbp]
    \centering
    \includegraphics[width=1\linewidth]{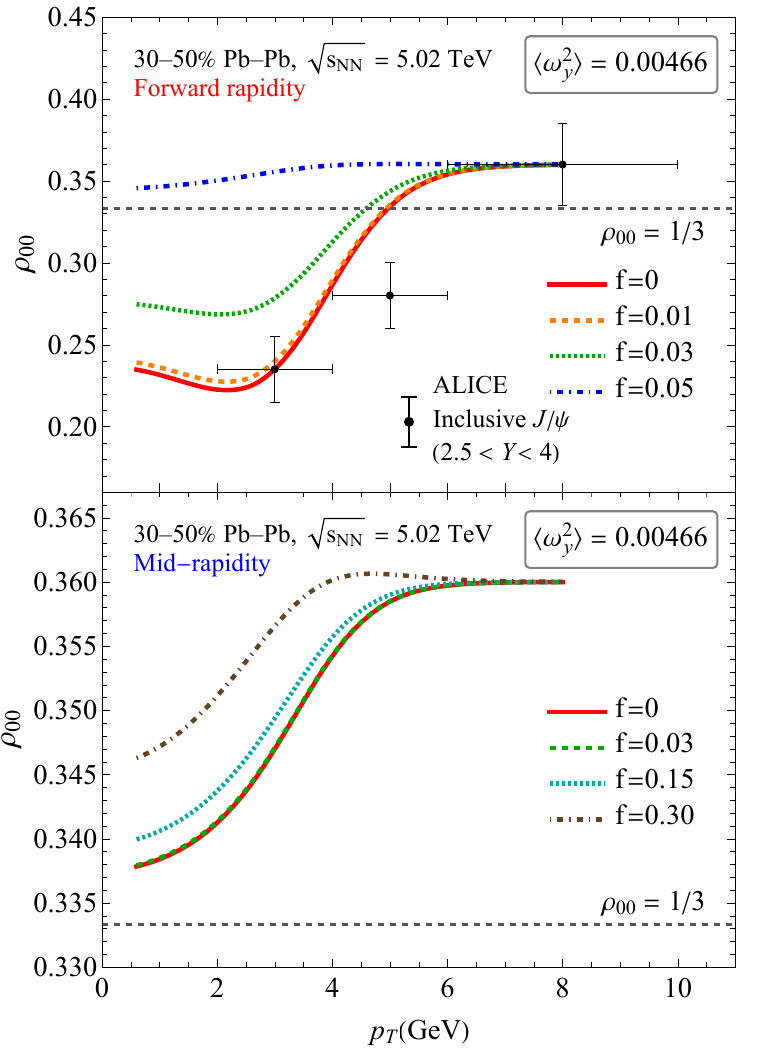}
    \caption{
    The $J/\psi$ spin density matrix element $\rho_{00}$ as a function of transverse momentum $p_T$ for different values of the isotropic vector-field fluctuation parameter $f$. The upper and lower panels show the results at forward rapidity and mid-rapidity, respectively, in Pb-Pb collisions at $\sqrt{s_{\mathrm{NN}}}=5.02$ TeV. The thermal-vorticity parameter is fixed to $\langle\omega_y^2\rangle=0.00466$. The horizontal black dashed lines denote the unpolarized reference value $\rho_{00}=1/3$.
}
    \label{withcolor}
\end{figure}

We now include the effect of the vector field. Thermal vorticity is not necessarily the only source of vector-meson spin alignment. In particular, the vorticity contribution is strongly suppressed at mid-rapidity, as discussed above. Therefore, a sizable positive deviation of $\rho_{00}$ from $1/3$ in this region may indicate the presence of additional spin-polarization mechanisms. The calculation with the vector field is based on the full expression in Eq.~(\ref{rho00boostnorm}). For simplicity, we assume an isotropic field fluctuation and set $f_T^2=f_z^2=f^2.$
The parameter $f$ is varied phenomenologically to illustrate the sensitivity of $\rho_{00}$ to the vector-field fluctuation strength. It should not be interpreted as a fitted parameter in the present analysis. For the vector-field study shown in Fig.~\ref{withcolor}, we fix the thermal-vorticity parameter to the upper boundary value $\langle\omega_y^2\rangle=0.00466$ as a representative input.

The numerical results including the vector-field contribution are shown in Fig.~\ref{withcolor}. The role of the vector field can be understood directly from Eq.~(\ref{rho00boostnorm}). The thermal-vorticity term appears with a sign that tends to decrease $\rho_{00}$ below $1/3$, whereas the vector-field fluctuation terms enter with the opposite tendency and increase $\rho_{00}$ in the present model. Therefore, at forward rapidity, the vector field partially compensates the negative deviation generated by the thermal vorticity in the coalescence component. As the value of $f$ increases, the low-$p_T$ result is pushed upward, and for sufficiently large $f$ the total $\rho_{00}$ can exceed the unpolarized reference value.

At mid-rapidity, the situation is different. Since the thermal-vorticity contribution is kinematically suppressed, the spin alignment generated by the vector field is less masked by the vorticity background. The vector field therefore provides a possible source of a positive low-$p_T$ deviation of $\rho_{00}$ from $1/3$. In the lower panel of Fig.~\ref{withcolor}, increasing $f$ enhances $\rho_{00}$ most clearly in the low and intermediate $p_T$ regions, where the coalescence component is still relevant. At high $p_T$, all curves approach the same limiting value $\rho_{00}^{\mathrm{init}}=0.36$. This is because the initial-production component dominates and the coalescence contribution becomes strongly diluted.

The values of $f$ used in the two rapidity regions are different in order to display the sensitivity of $\rho_{00}$ in each kinematic window. In the forward-rapidity panel, visible modifications already appear for $f=0.01$--$0.05$, while in the mid-rapidity panel larger values, up to $f=0.30$, are shown. This comparison illustrates the rapidity-dependent kinematic response of the spin-alignment observable to the underlying polarization sources. Thus, the low-$p_T$ behavior in the two rapidity regions is sensitive to different dominant mechanisms: at forward rapidity, it is mainly shaped by the competition between thermal-vorticity-induced coalescence and the initial-production baseline, whereas at mid-rapidity the vector-field fluctuation can become a more direct source of a positive low-$p_T$ enhancement.

\section{Summary and discussion\label{sec5}}

In this work, we have applied the relativistic spin Boltzmann framework for vector mesons to investigate the spin alignment of $J/\psi$ mesons in heavy-ion collisions. Within this formalism, we derived the non-relativistic expression for the spin density matrix element $\rho_{00}$ of heavy quarkonia. To interpret the experimental data, a two-component transport model was employed, wherein the observable $\rho_{00}^{\mathrm{obs}}$ is formulated as a weighted sum of the coalescence and primordial production components. We first evaluated the scenario governed solely by thermal vorticity, in the absence of vector-field contributions. In this case, the spin alignment of the coalescence component is driven entirely by the vorticity-induced polarization of charm quarks. At forward rapidity, the calculated $\rho_{00}$ exhibits a non-monotonic $p_T$ dependence, where the coalescence process dominates at low and intermediate $p_T$, while primordial production becomes dominant at high $p_T$. 

Furthermore, we provided predictions for the $J/\psi$ spin alignment at mid-rapidity based on this mechanism. We find that the thermal-vorticity contribution is strongly suppressed at mid-rapidity, a suppression that originates inherently from the Lorentz transformation between the $J/\psi$ rest frame and the laboratory frame. Consequently, the same thermal-vorticity source manifests as distinct observable effects across different rapidity windows. We further investigated the impact of effective vector-field fluctuations, introduced phenomenologically as a color-singlet Abelian-like field to capture local strong-interaction fluctuations. In our model, the vector field tends to enhance $\rho_{00}$. At forward rapidity, it partially compensates for the negative deviation induced by thermal vorticity, whereas at mid-rapidity—where vorticity is suppressed—it provides a direct source for the positive enhancement observed at low $p_T$. 

In summary, our results demonstrate that $J/\psi$ spin alignment is governed by an interplay of multiple physical mechanisms. The $p_T$ dependence of $\rho_{00}$ is intimately linked to the two-component production mechanism and the underlying charm-quark polarization, while the rapidity dependence is modulated by Lorentz transformations. Crucially, thermal vorticity and vector-field fluctuations contribute distinct, competing features to the coalescence component. The present results are obtained by retaining only the charm- and anticharm-quark polarization components along the global OAM direction. Polarization components along other directions may provide additional contributions to $J/\psi$ spin alignment and merit further investigation. A systematic comparison between forward- and mid-rapidity measurements will therefore serve as a powerful tool to disentangle these intertwined effects.

\vspace{1cm}
\appendix {\bf Acknowledgement:} 
This work is supported by the
National Natural Science Foundation of China (NSFC)
under Grant Nos. 12575149 and 12175165.

\bibliographystyle{apsrev4-2}
\bibliography{CompleteTransportModel}

\end{document}